\documentclass[12pt]{iopart}
\pdfoutput=1
\usepackage{cite}
\usepackage{iopams,graphicx}  
\usepackage{color}

\usepackage{stmaryrd}
\usepackage{makeidx}
\usepackage{indentfirst}
\usepackage{comment}

\begin{document}

\title[Reversible Von Neumann's expanding model]{One way to grow, many ways to shrink: the reversible Von Neumann expanding model}

\author{A. De Martino$\dag\ddag$, M. Figliuzzi$\ddag$, M. Marsili$\S$}

\address{$\dag$ CNR/IPCF and $\ddag$Dipartimento di Fisica, Sapienza Universit\`a di Roma, p.le A. Moro 2, 00185 Roma (Italy)}
\address{$\S$The Abdus Salam ICTP, Strada Costiera 14, 34014 Trieste (Italy)}
\ead{andrea.demartino@roma1.infn.it}
\begin{abstract}
We study the solutions of Von Neumann's expanding model with reversible processes for an infinite reaction network. We show that, contrary to the irreversible case, the solution space need not be convex in contracting phases (i.e. phases where the concentrations of reagents necessarily decrease over time). At optimality, this implies that, while multiple dynamical paths of global contraction exist, optimal expansion is achieved by a unique time evolution of reaction fluxes. This scenario is investigated in a statistical mechanics framework by a replica symmetric theory. The transition from a non-convex to a convex solution space, which turns out to be well described by a phenomenological order parameter (the fraction of unused reversible reactions) is analyzed numerically. 
\end{abstract}

\maketitle

\section{Introduction}

At the most basic level of abstraction, a transformation process like an industrial production technology or a chemical reaction can be specified merely by the coefficients measuring the amounts of each commodity or chemical species that are consumed and, respectively, produced when the process operates at, say, unit scale (or flux). When $N$ such processes are brought together in a network connecting $M$ species so that the inputs of one process may be the outputs of another, the emergent properties of the network  will depend crucially on the particular choice of the (quenched) ``stoichiometric'' coefficients. Of special interest in this respect are questions regarding network optimality. In a landmark paper \cite{JVN}, J. Von Neumann  considered the problem of finding the largest (uniform) species production rate possible for irreversible processes specified by given stoichiometry. The existence of a maximal expansion rate $\rho^\star$ can be proven easily for any matrices of input-output stoichiometric coefficients satisfying broad generic assumptions \cite{GALE}. 

The study of this problem with random stoichiometry \cite{a1,a2} provides hints on the typical behavior of the solution of the Von Neumann problem in complex situations. In the limit where the number ($N$) of reactions and ($M$) of species diverge, with a fixed ratio ($n=N/M$) a full characterization of the growth properties is possible, using tools of statistical physics. The maximal growth rate is given, to a first approximation, by the ratio of output and input coefficients, with a non-trivial correction which is of the order of $1/\sqrt{K}$, where $K$ is the (average) number of reactions in which each species is involved. This latter term depends on the structure of the network of reactions. In particular, as the ratio of processes to species increases, the maximal growth rate also increases. In addition, one can show that the space of solutions for a given growth rate is convex and that it shrinks to a single point when the growth rate approaches the maximal one, for a specific network of reactions  \cite{a1,a2}.

When the network of reactions is constrained to obey the mass conservation implicit in chemical reactions \cite{a3}, expanding solutions are not possible, as the maximal growth rate cannot be larger than zero. The Von Neumann problem then reduces to that of metabolite producibility in metabolic networks, addressed in \cite{IM2,IM}. The analysis of fluxes in the metabolic network of the bacterium {\it E. coli} within Von Neumann's framework has been discussed in \cite{PNAS}. Interestingly, the presence of conserved pools of metabolites imply that the solution space for the maximal growth rate does not coincide with a single point. Rather, a finite volume of flux configurations corresponding to maximal growth rate (equal to zero) exists. We refer the interested reader to \cite{PNAS} for the biological implications. Here we merely point out that even in this case the space of solutions retains its convexity properties: any linear combination of two solutions with positive weights is still a solution. 

These results apply to a network of irreversible reactions (or generically when reversible processes can be split into two separate reactions). This paper addresses the case where one or more of the reactions can also be run in the reverse direction. Such a scenario is not usually considered in economic applications, as technologies are generally unidirectional. For biochemical systems it is however crucial, both because many of the chemical reactions occurring in cells are physiologically reversible and because the shape of the solution space, particularly its convexity properties, has strong implications on the effectiveness of algorithms designed to find the optimal flux vectors and sample the solution space uniformly. 

We first show how the case in which processes can be reversed is a non-trivial generalization of the irreversible case. In particular, convexity is no longer guaranteed when the growth rate is negative, whereas it can still be proven when the growth rate is non-negative. In order to address what happens in typical cases, we replicate the analysis on random systems. Reversibility turns out to generate a substantially more complex scenario for the solution space of Von Neumann's expansion problem, which considerably complicates its numerical analysis. For simplicity, we focus on the fully-connected version, defined in Sec. 2, resorting to the replica trick to compute the maximal growth rate (Sec. 3). In Section 4 the analytic solution is discussed and compared with numerical results obtained for a large network with a given stoichiometry. This analysis confirms that, when the growth rate is negative, the space of solutions splits into many disjoint components. A brief outlook is given in the concluding Section 5. Finally, an Appendix details the heuristic algorithm employed for the numerical exploration of the solution space.

\section{The problem and a basic observation}

Let us consider $N$ processes, labeled by $i=1,\ldots,N$, which operate on a set of $M$ species -- be them chemical substances or commodities -- labeled by $\mu=1,\ldots,M$. These are transformation processes whereby some inputs are transformed into some output products in proportions which are specified by the stoichiometric matrices of outputs $\mathbf{A}=\{a_i^\mu\}$ and inputs $\mathbf{B}=\{b_i^\mu\}$. Time is discrete, and we consider the case where all processes are run in parallel at scale $\sigma_i(t)\geq 0$, between time $t$ and time $t+1$. This means that $\sigma_i(t)$ processes of type $i$ are run in parallel, consuming $\sigma_i(t) b_i^\mu$ units of species $\mu=1,\ldots,M$ and producing 
$\sigma_i(t) a_i^{\mu'}$ units of species $\mu'=1,\ldots,M$.

Hence, the total quantity of species $\mu$ consumed as input at time $t$, by all processes, is $I^\mu(t)=\sum_{i=1}^N \sigma_i(t)b_i^\mu$ and the total amount of output of species $\mu$ produced is 
$O^\mu(t)=\sum_{i=1}^N \sigma_i(t)a_i^\mu$. Given an initial concentration $O^\mu(0)$ of species, feasible growth paths $\{\boldsymbol{\sigma}(t)\}_{t> 0}$ (with $\boldsymbol{\sigma}(t)=\{\sigma_i(t)\}_{i=1}^N$) are those such that $\sigma_i(t)\ge 0$ for all $i$ and $t$, and that the process at time $t$ produces enough outputs to run the processes at time $t+1$, i.e. $O^\mu(t)\ge I^\mu(t+1)$ for all $t$. Von Neumann \cite{JVN} further focuses on paths $\sigma_i(t)=s_i \rho^t$, with $s_i\geq 0$, of exponential expansion ($\rho>1$) or contraction ($\rho<1$). Feasible solutions $\mathbf{s}$ with growth factor $\rho$ (and growth rate $\log\rho$) are easily seen to satisfy the constraints
\begin{equation}
\label{vndiseq}
c^\mu(\mathbf{s})\equiv\sum_{i=1}^N s_i \left(a_i^\mu-\rho b_i^\mu\right)\ge 0,\qquad\forall \mu=1,\ldots,M
\end{equation}
or, more compactly, $(\mathbf{A}-\rho\mathbf{B})\mathbf{s}\geq \mathbf{0}$. If $\rho$ is very small, the space of solutions is very large, in particular any $\mathbf{s}\ge \mathbf{0}$ is a solution for $\rho=0$. On the contrary, if $\rho$ is large enough no vector $\mathbf{s}$ will satisfy the conditions (\ref{vndiseq}). For any $\mathbf{A}$ and $\mathbf{B}$ satisfying broad generic assumptions \cite{GALE} it is easy to show that there is a maximal expansion rate $\rho^\star$. Von Neumann's (VN) problem amounts to finding the maximal growth rate and the path of maximal growth, i.e. 
\begin{equation}
\max_{\mathbf{s}\ge\mathbf{0}}\rho~~~~~{\rm subject~to}~\left(\mathbf{A}-\rho\mathbf{B}\right)\mathbf{s}\geq \mathbf{0}.
\end{equation}

Let us now generalize this problem to introduce reversible processes. 
Reversibility implies that input and output coefficients can be interchanged. If process $i$ is reversible, it is possible to produce a quantity $b_i^\mu$ of species $\mu$ by consuming $a_i^{\mu'}$ units of species $\mu'$. It is interesting to observe at the outset that reversing the direction of reactions does not correspond to merely reversing the direction of time. Indeed, imagine that a solution $\mathbf{s}$ with growth rate $\rho$ exists. Then one would na\"ively expect that, if all reactions are reversed, the situation where all processes are run in the same proportions (i.e. with operation scales $\bar{\mathbf{s}}=\mathbf{s}$) should yield a solution with $\bar\rho=1/\rho$. Indeed, it is not difficult to find that the maximal growth rate $\bar\rho$ corresponding to the reversed solution satisfies
\begin{equation}
\label{trev}
\bar\rho\le \frac{1}{\rho}\min_\mu\left(1-\frac{c^\mu(\mathbf{s})}{O^\mu(\mathbf{s})}\right)
~~~,~~~~~O^\mu(\mathbf{s})=\sum_{i=1}^N s_i a_i^\mu
\end{equation}
which means that in order to obtain a growth rate $\bar\rho=1/\rho$ for the reversed process the condition $c^\mu=0$ for all $\mu$ is necessary but not sufficient\footnote{In order to show this, let $\bar c^\mu(\mathbf{s})=\sum_i s_i(b_i^\mu-\bar\rho a_i^\mu)$ and note that
\begin{equation*}
c^\mu(\mathbf{s})+\rho\bar c^\mu (\mathbf{s}) = (1-\rho\bar\rho)O^\mu(\mathbf{s})
\end{equation*}
This immediately tells us that if $c^\mu(\mathbf{s})=\bar c^\mu (\mathbf{s})=0$ for all $\mu$, then $\bar\rho=1/\rho$. If however $\bar c^\mu(\mathbf{s})\geq 0$, one has $c^\mu(\mathbf{s})\leq (1-\rho\bar\rho) O^\mu(\mathbf{s})$. Eq. (\ref{trev}) is derived straightforwardly from this.}. 
Hence reversing reactions is not, in general, equivalent to time reversal.

By the same arguments as above, one finds that, when reactions can be run in both directions, Von Neumann's (VN) problem can be cast as
\begin{equation}\label{vn}
\max_{\mathbf{s}}\rho~~~~~{\rm subject~to}~\mathbf{\Xi}_\rho(\mathbf{s})\mathbf{s}\geq\mathbf{0}
\end{equation}
where $\mathbf{s}$ is a flux vector and $\mathbf{\Xi}_\rho(\mathbf{s})$ is a $M\times N$ matrix with entries given by
\begin{equation}
\xi_i^\mu(\rho,s_i)=
\cases{a_i^\mu-\rho b_i^\mu&{\rm for $s_i> 0$}\\
b_i^\mu-\rho a_i^\mu&{\rm for $s_i< 0$}}
\end{equation}
Clearly, because the matrices depends on scales $\mathbf{s}$, the above conditions are not linear in $\mathbf{s}$. This generates key differences between the irreversible case and the present model.  In the former case, in fact, a convex combination $\lambda\mathbf{s}+\lambda'\mathbf{s'}$ of two solutions $\mathbf{s}$ and $\mathbf{s'}$ with non negative coefficients $\lambda$ and $\lambda'$ is always a feasible solution of the original problem. Irreversibility can modify this picture. To see this it is convenient to introduce the quantities
\begin{equation}
c^\mu(\mathbf{s})\equiv\sum_{i=1}^N s_i [\theta(s_i)(a_i^\mu-\rho b_i^\mu)-\theta(-s_i)(b_i^\mu-\rho a_i^\mu)] 
\end{equation}
which are required to be non-negative for each $\mu$ under conditions (\ref{vn}). Indeed substituting one finds 
\begin{equation}
\label{eqZ}
c^\mu(\lambda\mathbf{s}+\lambda'\mathbf{s'})=\lambda c^\mu(\mathbf{s})+\lambda' c^\mu(\mathbf{s'})+(1-\rho)\mathcal{A}^\mu(\mathbf{s},\mathbf{s'})
\end{equation}
where
\begin{equation}\label{eqA}
\fl
\mathcal{A}^\mu(\mathbf{s},\mathbf{s'})=\sum_i(a_i^\mu+b_i^\mu)[\lambda s_i\theta(-s_i)+\lambda' s_i'\theta(-s_i')-(\lambda s_i+\lambda' s_i')\theta(-\lambda s_i-\lambda' s_i')]
\end{equation}
If $\mathbf{s}$ and $\mathbf{s'}$ are both solutions, then a sufficient condition for their linear combination to be again a solution is that $(1-\rho)\mathcal{A}^\mu(\mathbf{s},\mathbf{s'})\geq 0$. If $s_i$ and $s_i'$ have the same sign, for all $i$ (i.e. $s_i s_i'>0$), then $\mathcal{A}^\mu(\mathbf{s},\mathbf{s'})=0$, which means that convexity is guaranteed for all values of $\rho$. But if the two solutions differ by a change of sign (i.e. if $s_i s_i'<0$ for some $i$), then the term in square brackets in Eq. (\ref{eqA}) is non positive and it equals $-\min[\lambda |s_i|,\lambda' |s_i'|]\le 0$. This implies that convexity (i.e. $(1-\rho)\mathcal{A}^\mu(\mathbf{s},\mathbf{s'})\geq 0$) is guaranteed only if $\rho\geq 1$. Note that convexity may not occur even at $\rho^\star$ when $\rho^\star<1$. 

Hence when $\rho<1$ it is possible that the space of feasible solutions is disjoint in different connected pieces. However the condition  $(1-\rho)\mathcal{A}^\mu(\mathbf{s},\mathbf{s'})\geq 0$ for all $\mu$ is sufficient but not necessary, as $c^\mu(\lambda\mathbf{s}+\lambda'\mathbf{s'})\ge 0$ may be satisfied for all $\mu$, even if the former condition is not satisfied. A trivial case when this happens is the case $\rho=0$, where clearly all $\mathbf{s}\in \mathbb{R}^N$ are solutions. Hence, in order to understand whether non-convexity is  realized in typical cases, we're going to study ensembles of random instances in the next section.

\section{Typical properties of feasible solutions: Replica-symmetric theory}

In what follows we are going to take 
\begin{eqnarray}
a_i^\mu=\bar a\left(1+\frac{\alpha_i^\mu}{\sqrt{N}}\right)\\
b_i^\mu=\bar b\left(1+\frac{\beta_i^\mu}{\sqrt{N}}\right)
\end{eqnarray}
where $\alpha_i^\mu$ and $\beta_i^\mu$ are going to be drawn, independently from some distribution with finite first and second moments. We assume that $\phi N$ reactions are reversible and $(1-\phi)N$ are not ($0\leq\phi\leq 1$).  To leading order in $N$ we have $\rho^\star=\max\{\frac{\bar a}{\bar b},\frac{\bar b}{\bar a}\}$ if $\phi> 0$, which can be achieved by running the reversible reactions in the maximal growth direction while switching off ($s_i=0$) all reactions with growth rate $\min\{\frac{\bar a}{\bar b},\frac{\bar b}{\bar a}\}$. The nontrivial part of the growth rate is hence related to terms of order $1/\sqrt{N}$ and, in order to focus on those, we shall set $\bar a=\bar b$. As in the irreversible case \cite{a1}, to account for subleading effects we set
\begin{equation}
\rho=1+\frac{g}{\sqrt{N}}
\end{equation} 
and focus on $g$. The maximum allowed $g$ shall be denoted by $g^\star$.

\subsection{Replica theory}

When $N,M\to\infty$ with $n=N/M$ fixed, one can apply the replica trick to calculate $g^\star$ (which is expected to be a self-averaging quantity). We can argue as in \cite{a1}: the volume of solutions for a specific choice of the stoichiometry is
\begin{equation}
V(\rho)={\rm Tr}_{\mathbf{s}}\prod_{\mu=1}^M\theta[c^\mu(\mathbf{s})]\delta\left(\sum_{i=1}^N|s_i|-N\right)
\end{equation}
where the trace over reversible reactions involves integrals from $-\infty$ to $+\infty$, that over irreversible ones from $0$ to $+\infty$. The $\delta$-function enforces the constraint $\sum_i|s_i|=N$, which sets a scale for the fluxes. In the thermodynamic limit one expects the typical volume of solutions to be given by $\mathcal{V}\propto \exp(Nv(\rho))$ where
\begin{equation}
v(\rho)=\lim_{N\to\infty}\frac{1}{N}\overline{\log V(\rho)}~~~,
\end{equation}
the over-bar denoting an average over the quenched disorder $\{\alpha_i^\mu,\beta_i^\mu \}$. By the replica trick
\begin{equation}
v(\rho)=\lim_{N\to\infty}\lim_{r\to 0}\frac{1}{Nr}\log\overline{V(\rho)^r}
\end{equation}
After expressing the $\theta$ function via its Fourier decomposition, carrying out the disorder average and isolating the emergent order parameter
\begin{equation}
q_{\ell\ell'}=\frac{1}{N}\sum_{i=1}^N s_{i\ell}s_{i\ell'}
\end{equation}
one arrives at
\begin{equation}
\overline{V(\rho)^r}=\int J_1(\mathbf{q})J_2(\mathbf{q})d\mathbf{q}
\end{equation}
\begin{equation}
J_1(\mathbf{q})=\int_{-\infty}^\infty D\mathbf{z}\int_0^\infty D\mathbf{c}
\prod_{\mu=1}^M\exp\left[{\rm i}\sum_\ell z_\ell^\mu(c^\mu_\ell+g)-\frac{k}{2}\sum_{\ell,\ell'}q_{\ell\ell'}
z_\ell^\mu z_\ell'^\mu\right]
\end{equation}
\begin{equation}\label{j2}
J_2(\mathbf{q})={\rm Tr}_{\mathbf{s}}\prod_\ell\delta\left(\sum_{i=1}^N|s_{i\ell}|-N\right)\prod_{\ell\leq\ell'}
\delta\left(\sum_{i=1}^N s_{i\ell}s_{i\ell'}-Nq_{\ell\ell'}\right)
\end{equation}
where $\ell,\ell'$ run from $1$ to $r$ and $k=\overline{(\alpha_i^\mu-\beta_i^\mu)^2}$. 

Under the replica symmetric (RS) Ansatz where
\begin{equation}
q_{\ell\ell'}=q+\chi\delta_{\ell\ell'}
\end{equation}
one finds that at the relevant saddle point
\begin{equation}
v(g)=\max_{q,\chi}[H_1+\max_{m,\beta,\tau}H_2]\label{vg}
\end{equation}
where $m, \beta$ and $\tau$ are additional order parameters, derived from imposing RS Ans\"atze on the Lagrange multipliers enforcing the $\delta$-functions in (\ref{j2}), and
\begin{eqnarray}
H_1=&\frac{1}{n}\mathbb{E}_\xi\log\int_0^\infty\frac{dc}{\sqrt{2\pi k\chi}}\exp\left[-\frac{(c+g+\xi\sqrt{kq})^2}{2k\chi}\right]\\
H_2=&m+\frac{\beta}{2}(q+\chi)-\frac{\tau^2}{2}\chi+(1-\phi)\mathbb{E}_\xi
\log\int_0^\infty ds \exp\left[-\frac{\beta}{2}s^2-(m+\tau\xi)s\right]\nonumber\\
&+\phi\mathbb{E}_\xi\log\int_{-\infty}^\infty ds \exp\left[-\frac{\beta}{2}s^2-m|s|-\tau s\xi\right]
\end{eqnarray}
where 
\begin{equation}
\mathbb{E}_\xi\cdots=\frac{1}{\sqrt{2\pi}}\int\cdots e^{-\xi^2/2}d\xi
\end{equation}

First note that for $k=0$ the problem has the trivial solution $g=0$. Indeed, in the limit $k\to 0$ the integral in $H_1$ is dominated by $c=0$, and its value is largest for $g=0$. So it is precisely the presence of fluctuations in input-output coefficients that allows for expanding phases or imposes contracting phases. 

The last two terms in $H_2$ contain contributions of reversible and irreversible processes, respectively.
The last term arises from the integration on the variable $s_i\in\mathbb{R}$ of a {\em representative} reversible process, whereas in the last but first term the integration is limited on $s_i\ge 0$, as appropriate for reversible processes. The problem in Eq. (\ref{vg}) has to be solved numerically for each values of the parameters.

\subsection{Optimal growth solution}

The solution simplifies in the limit $\rho\to\rho^\star$ (or $g\to g^\star$) of maximal growth rate.
Indeed, assuming that a unique flux vector survives in this limit, amounts to studying the solutions in the limit $\chi\to 0$, since $\chi$ is proportional to the Euclidean distance between two solutions $\mathbf{s}_\ell$ and $\mathbf{s}_{\ell'}$:
\begin{equation}
\frac{1}{N}\sum_i(s_{i\ell}-s_{i\ell'})^2=2\chi~~~~~(\ell\neq\ell')
\end{equation}
Evaluating $H_1$ in this limit one finds that the integral is dominated by values of the constraints $c$ given by
\begin{equation}
c^\star(\xi)=
\cases{-g^\star-\xi\sqrt{kq}&{\rm for $\xi<-g^\star/\sqrt{kq}$}\\
0&{\rm otherwise}}
\end{equation}
so that for the distribution of $c$ one finds
\begin{equation}
p(c)=\frac{e^{-\frac{(c+g^\star)^2}{2qk}}}{\sqrt{2\pi qk}}\theta(c)+c_0\delta(c)
~~~~~,~~~~~c_0=\frac{1}{2}+\frac{1}{2}{\rm erf}\frac{g^\star}{\sqrt{2kq}}.
\end{equation}
For fluxes, it is convenient to study reversible and irreversible processes separately. In order to extrapolate the dominant contribution in the limit $\chi\to 0$, we set
\begin{equation}
b=\beta\chi~~~,~~~t=\tau\chi~~~,~~~tz=-m\chi.
\end{equation}
In the case of irreversible fluxes, one has
\begin{equation}
s^\star_{irr}(\xi)=
\cases{t(z-\xi)/b&{\rm for $\xi<z$}\\0&{\rm otherwise}}
\end{equation}
and the corresponding distribution of fluxes is given by
\begin{equation}
p_{irr}(s)=\psi_{0,irr} \delta(s)+\theta(s)\frac{b e^{-\frac{(s-tz/b)^2}{2(t/b)^2}}}{t\sqrt{2\pi}}
~~~~~,~~~~~\psi_{0,irr}=\frac{1}{2}{\rm erfc}\frac{z}{\sqrt{2}}
\end{equation}
Some greater care is required for reversible reactions. It turns out that, at the optimal growth $\rho^*$, the integral in $s$ for reversible fluxes is dominated by
\begin{equation}
s^\star_{rev}(\xi)=
\cases{t(z-\xi)/b&{\rm for $\xi<{\rm min}\{0,z\}$}\\
t(-z-\xi)/b&{\rm for $\xi>{\rm max}\{0,-z\}$}\\
0&{\rm otherwise}}.
\end{equation}
Note that for $z>0$ the support of the distribution of fluxes excludes the interval $[-tz/b,tz/b]$ around the origin. Therefore, in such conditions solutions with null reversible fluxes are forbidden, i.e. each reversible reaction is active either in one direction or the other. If $z<0$, instead, the support of the distribution extends over the whole real axis, but a finite fraction $\psi_{0,rev}$ of reversible fluxes are zero. Finally, one finds
\begin{eqnarray}
p_{rev}(s)=\psi_{0,rev}\delta(s)+\theta(z)f_+(z)+\theta(-z)f_-(z)
\end{eqnarray}
where
\begin{eqnarray}
f_+(z)=\theta(s)\frac{b e^{-\frac{(s-tz/b)^2}{2(t/b)^2}}}{t\sqrt{2\pi}}+\theta(-s)\frac{b e^{-\frac{(s+tz/b)^2}{2(t/b)^2}}}{t\sqrt{2\pi}}\\
f_-(z)=\theta(s+tz/b)\frac{b e^{-\frac{(s-tz/b)^2}{2(t/b)^2}}}{t\sqrt{2\pi}}+\theta(-s+tz/b)\frac{b e^{-\frac{(s+tz/b)^2}{2(t/b)^2}}}{t\sqrt{2\pi}}
\end{eqnarray}
while
\begin{equation}
\psi_{0,rev}={\rm Prob}\bigg\{\xi\in[\min\{0,z\},\max\{0,-z\}]\bigg\}
\end{equation}
is the probability that a reversible process is inactive (to which only negative values of $z$ contribute). Summing up, the distribution of fluxes is given by:
\begin{equation}
p(s)=\phi p_{rev}(s)+(1-\phi)p_{irr}(s)
\end{equation}
All of the above quantities can be evaluated directly once saddle point equations are solved for the order parameters and for $g^\star$ as a function of $n$ and $\phi$.

\section{Results}

\begin{figure}
\begin{center}
\includegraphics[width=9cm]{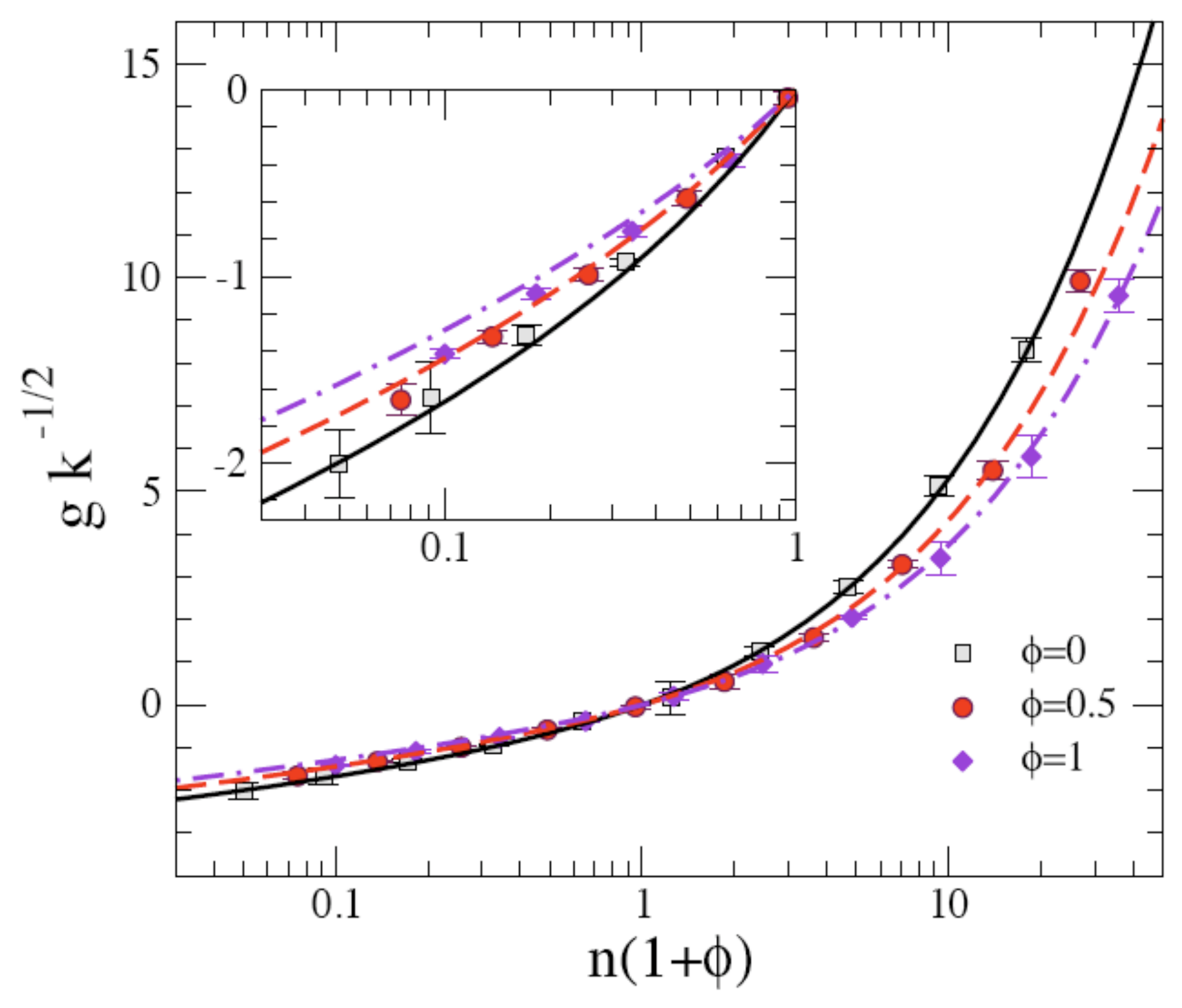}
\caption{\label{gstar}$g^\star/\sqrt{k}$ versus $n_{ef}=n(1+\phi)$. Lines represent analytic (replica) predictions, markers (with bars corresponding to the numerical error) results from simulations. Inset: detail for $n_{ef}<1$.}
\end{center}
\end{figure}

The numerical solution of the saddle point equations, in the optimal growth case, exhibits the following scenario.
An expanding phase with $g^\star>0$ is possible if $n>1/(1+\phi)$ or $n_{ef}\equiv n(1+\phi)>1$, while the system is necessarily confined to a contracting regime if $n_{ef}<1$ (see Fig. \ref{gstar}). This is in line with the results derived for the irreversible case, in view of the fact that $N(1+\phi)$ is the total number of processes available to the system. 
Increasing $\phi$, at constant $n_{ef}$, i.e. increasing the share of reversible processes, appears to have a beneficial effect in the contracting regime (larger $\phi$ means larger $g^\star$ ), while the opposite occurs in the expanding regime (see Fig. \ref{gstar}).

\begin{figure}
\begin{center}
\includegraphics[width=7.5cm]{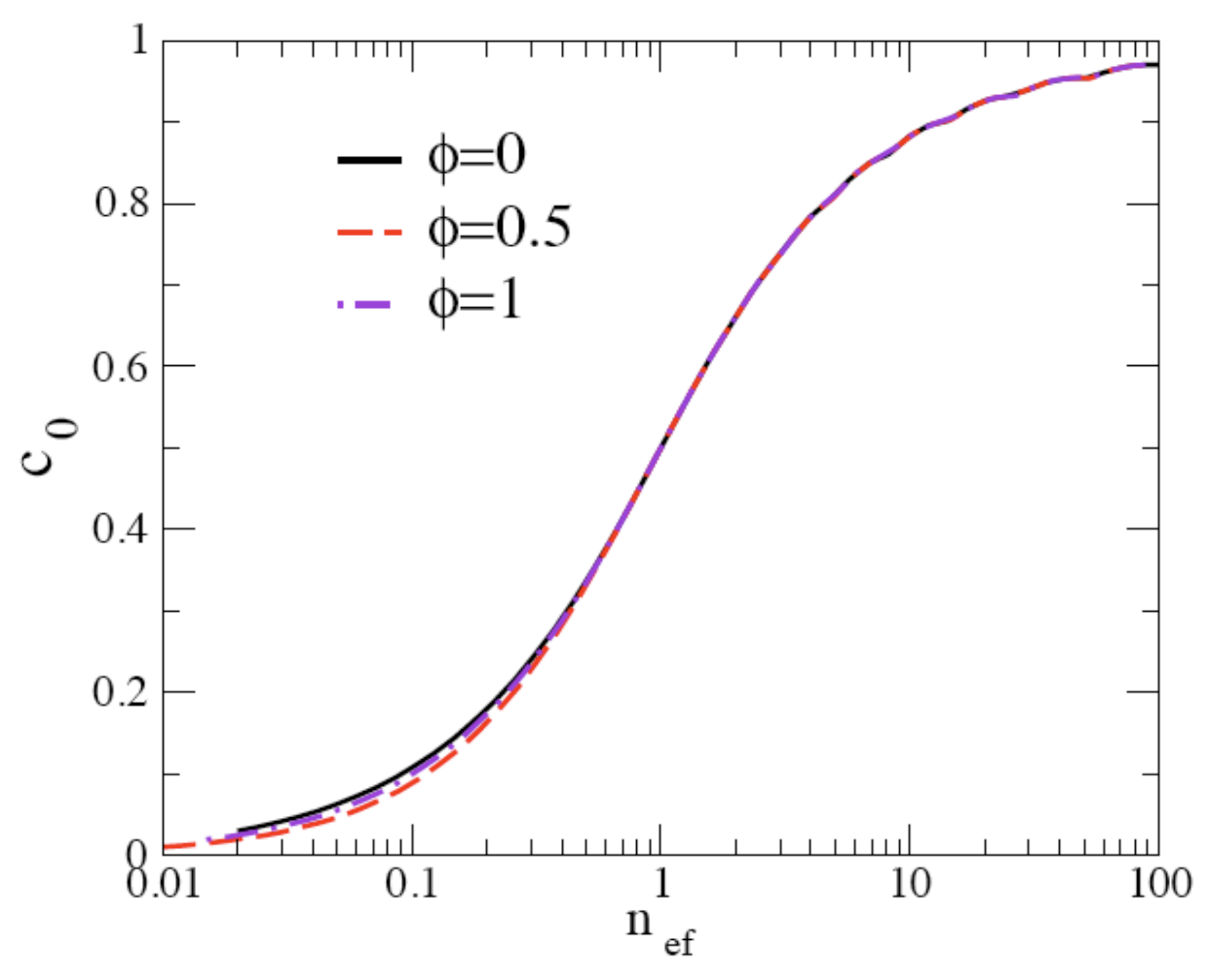}
\includegraphics[width=7.5cm]{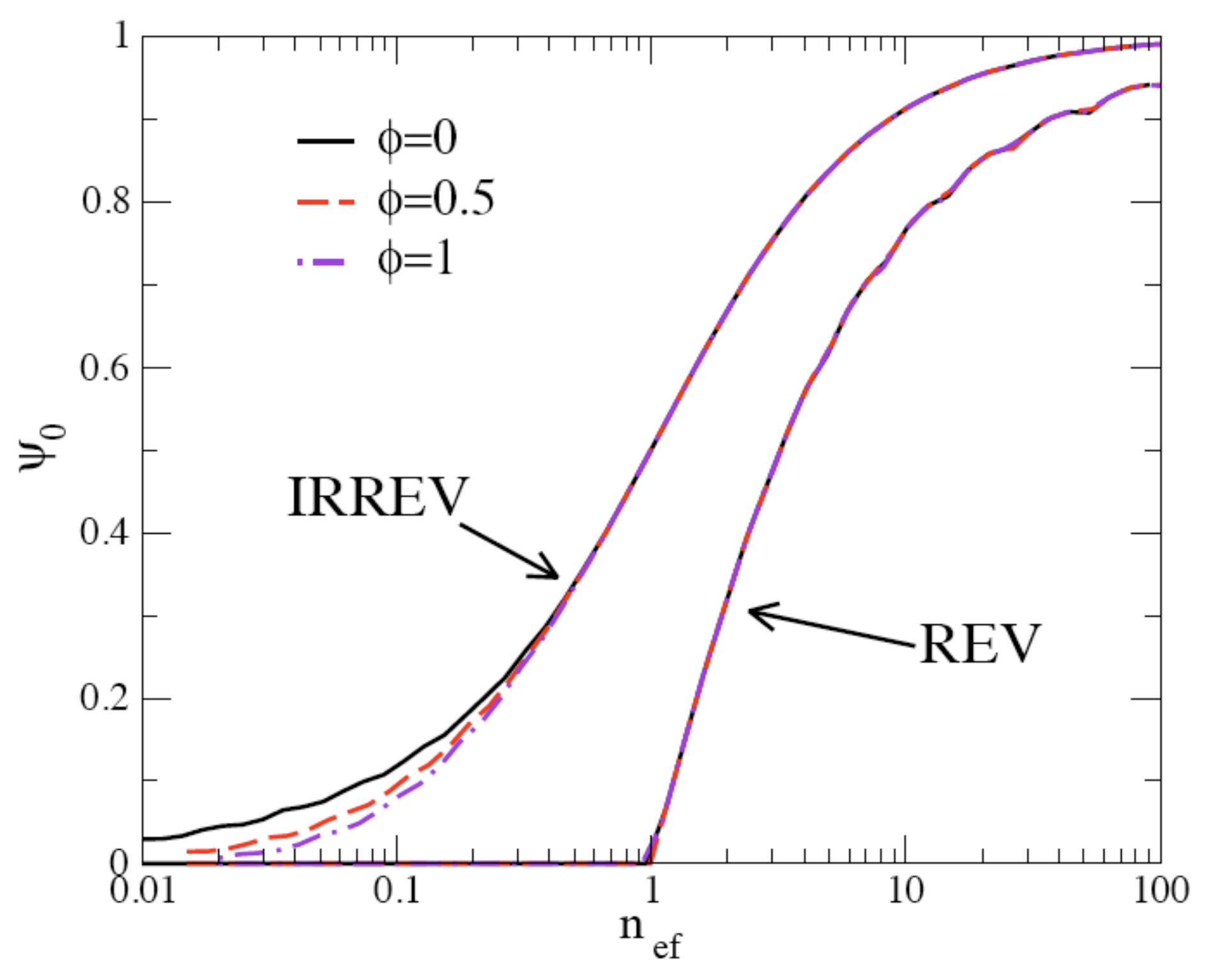}
\caption{\label{repl}Left panel: fraction of intermediate reagents ($c_0$) versus $n_{ef}$ for different values of $\phi$. Right panel: fraction of inactive processes ($\psi_0$) versus $n_{ef}$ for different values of $\phi$. The reversible and irreversible components are shown separately.}
\end{center}
\end{figure}

At $g^\star$, the fraction of reagents that are not produced (`intermediates', with $c^\mu=0$) behaves similarly to the purely irreversible model as a function of $n_{ef}$ (see Fig. \ref{repl}, left). 
In brief, as higher growth rates become achievable, the process becomes more efficient, with a decreasing fraction of ``wasted'' reagents ($c^\mu>0$). For the fractions of inactive processes one observes strikingly different behaviors for the reversible and irreversible components (Fig. \ref{repl}, right). The former, in particular, suggests the existence of a second-order phase transition at $n=1/(1+\phi)$: for $n_{ef}<1$ (i.e. in the contracting regime) all reversible processes are active, while $\psi_{0,rev}>0$ strictly in the expanding regime with $n_{ef}>1$.

\begin{figure}
\begin{center}
\includegraphics[width=7.5cm]{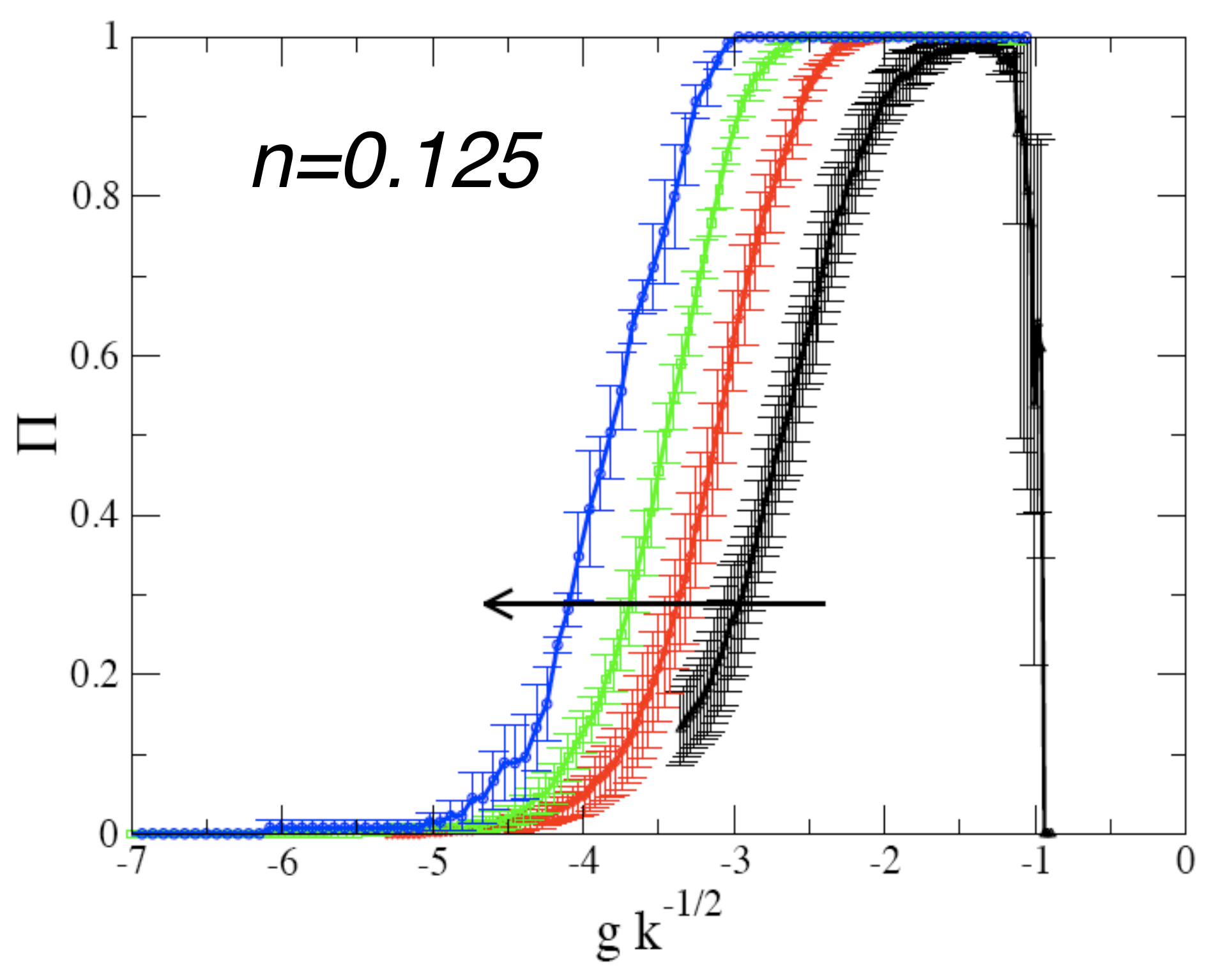}
\includegraphics[width=7.5cm]{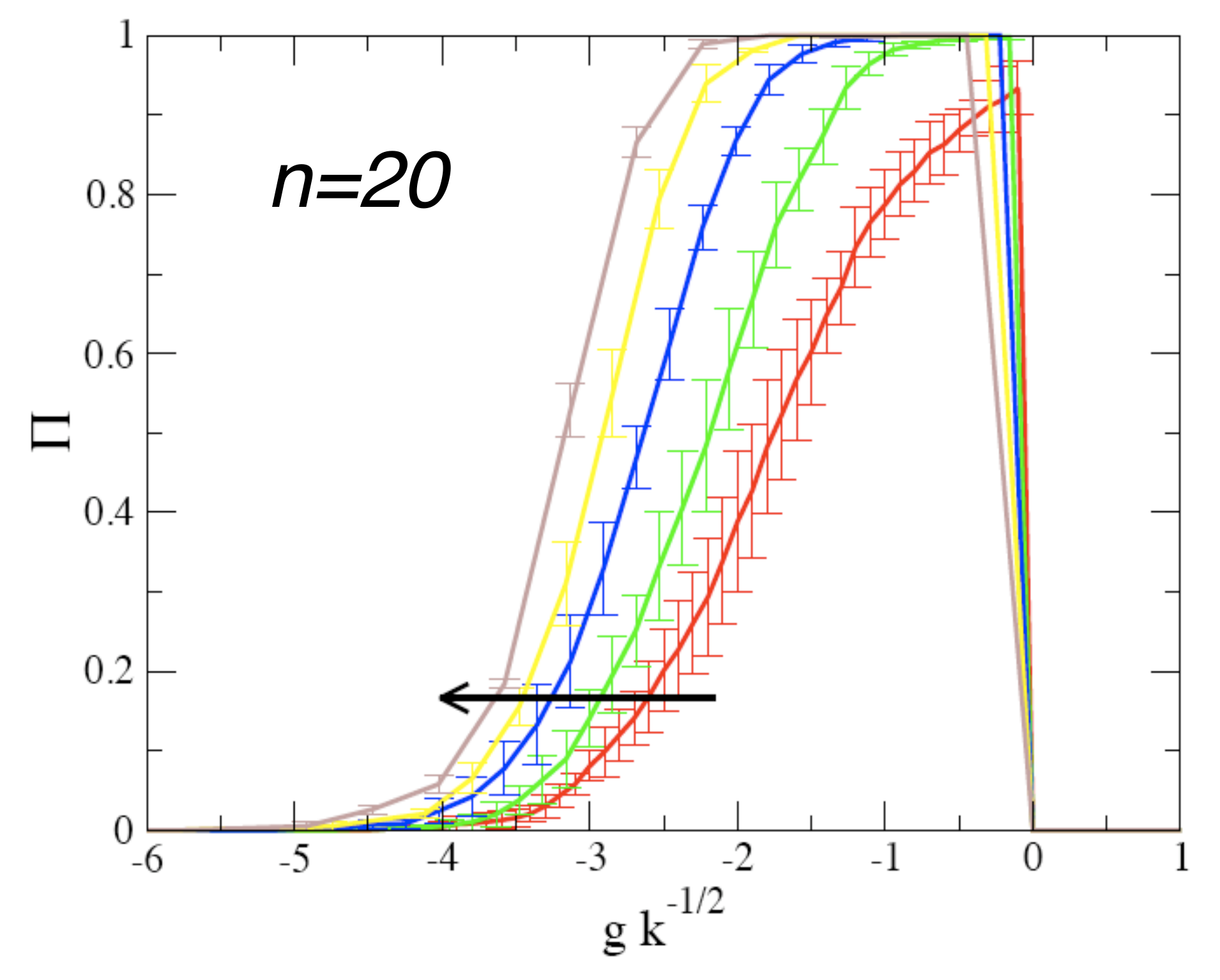}
\caption{\label{conv_n1}Numerical estimate of the probability $\Pi$ that the linear combination of two solutions with coefficients $1/2$ doesn't solve the fully reversible Von Neumann problem ($\phi=1$). Values of $n=N/M$ are as reported in the panels. Note that the optimal growth rate lies in the contracting (resp. expanding) phase in the left (resp. right) panel. System sizes are as follows: $N=10,25,50$ and $100$ increasing in the direction of the arrow for $n=1/8$ (note that data for $N=10$ may suffer from small system size); $N=200,500,1000,2000$ and $4000$ increasing in the direction of the arrow for $n=20$.}
\end{center}
\end{figure}

To verify these predictions we have computed optimal growth solutions numerically by an extension of the Minover$^+$ algorithm defined in \cite{a3} that accounts for reversible reactions. Details are given in the Appendix. It turns out that the analytic calculation is in very good agreement with numerical estimates in the expanding phase (see Fig. \ref{gstar}). In the contracting regime, however, one observes deviations from the analytic curves that increase with $\phi$. In the light of the previous discussion, it is reasonable to expect that such deviations are related to a breakdown of convexity at $g^\star$. To test this hypothesis, we have measured, at fixed $N$ and $M$, the probability $\Pi$ that the linear combination of two solutions with coefficients $1/2$ is not a solution of the original problem (see Fig. \ref{conv_n1}).
One observes clear signs of convexity breakdown for $g<0$ (or $\rho<1$), while the solution space appears to be convex both for $g>0$ and when $g$ is sufficiently smaller than $g^\star$. When the optimal solution lies in the contracting regime, $\Pi$ displays strong numerical fluctuations close to $g^\star$ making it hard to identify unambiguously whether a single solution survives. Note that, while the transition at $g=0$ is correctly located by such an analysis, the crossover point seen numerically for $g<0$ is just an upper bound of the real transition point where the solution space turns from being convex to non-convex upon increasing $g$.

\begin{figure}
\begin{center}
\includegraphics[width=9cm]{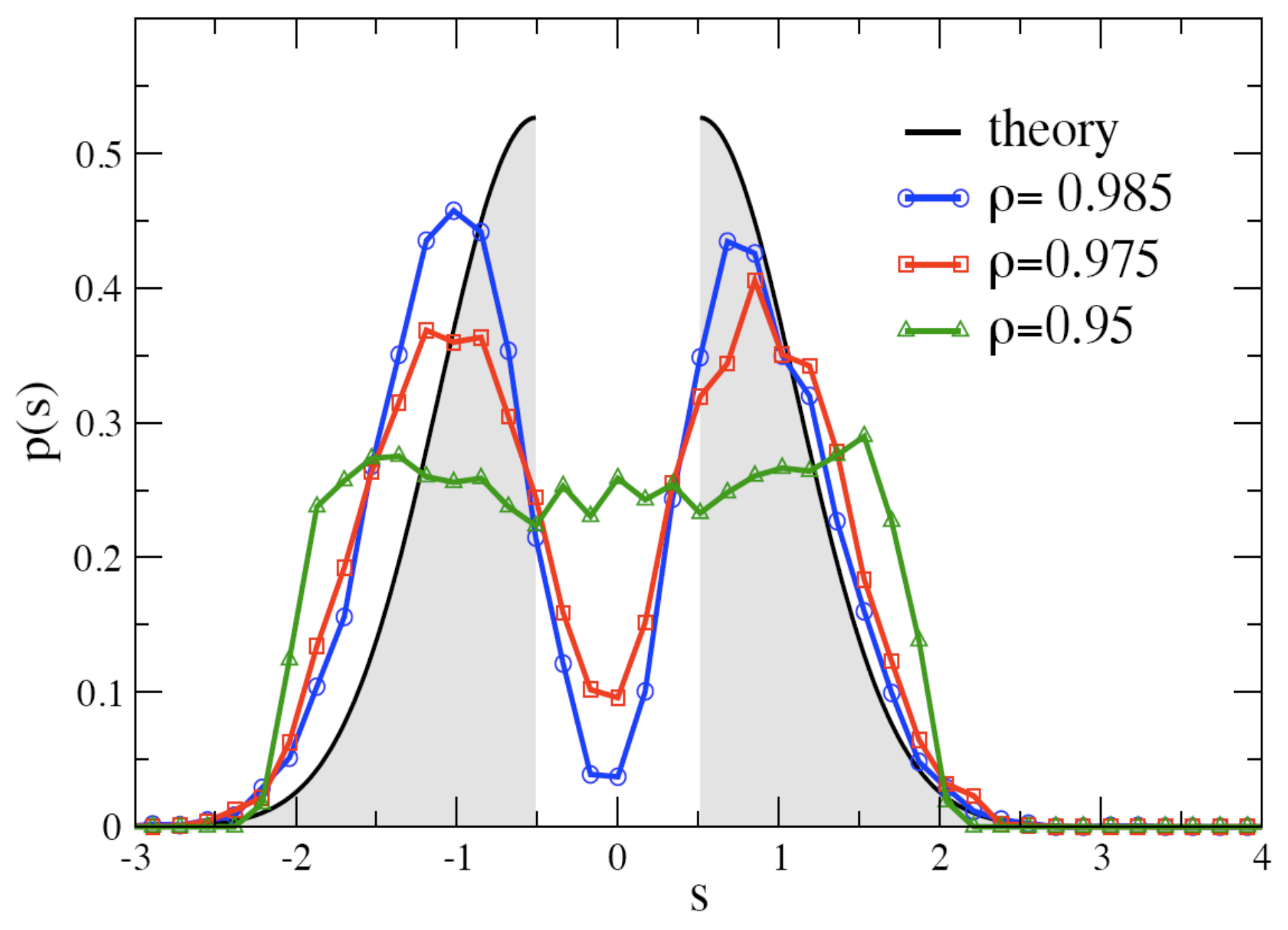}
\caption{\label{fluxdis}Distribution of $s$ for reversible processes for $n_{ef}=1/4$ (contracting phase) and $\phi=1$ ($M=8N=800$). The analytical prediction (continuous black line) is shown together with distributions for increasing values of $\rho<\rho^\star$. For these simulations, we estimate $\rho^\star\simeq 0.986$.}
\end{center}
\end{figure}

We have estimated numerically the flux distributions in the contracting phase for $\phi=1$ and $n_{ef}=1/4$ (see Fig. \ref{fluxdis}). 
The non convexity of the solution space makes it hard to achieve the optimal growth solutions (especially so for the negative part of the flux distribution). In spite of this, our results clearly show that as one gets closer to $g^\star$ the distribution of fluxes develops the double peak structure predicted by the replica theory. 

\section{Outlook}

Von Neumann's expanding model acquires considerable complexity in the presence of reversibility. Our results show in particular that multiple dynamical paths for reaction rates exist in contracting phases due to a breakdown of convexity of the solutions' space, whereas a convex solution space, with an unique optimal trajectory, characterizes expanding phases. Our theory is able to describe the expanding phase and does appear to capture some of the salient features of the contracting regime. At the same time, this study opens a number of interesting avenues for future research: In first place, it would be important to understand whether the non-convex solution space for $\rho<1$ can be described by a replica symmetry breaking Ansatz. Secondly, non-convexity poses the problem of designing efficient algorithms to sample the solution space in the contracting regime. The algorithm presented here is a promising first step in this direction.

\ack This work was partially supported by the IIT Seed Project DREAM.

\section*{Appendix. Algorithm to compute $\rho^\star$}

The Minover$^+$ algorithm \cite{a2} exploits the similarity between Von Neumann's conditions and the pattern storage condition in the perceptron to compute $\rho^\star$ on any graph in the irreversible Von Neumann model where fluxes are semipositive. In this case, the procedure to find a solution at a given $\rho$ is based on iteratively rotating the flux vector $\mathbf{s}$ in the direction of the least satisfied constraint, similar to \cite{KM}:
\begin{enumerate}
\item At step $j=0$: randomly initialize the flux vector $\mathbf{s}(0)$;
\item At step $j+1$: compute 
\begin{equation}\label{muz}
\mu_0={\rm arg}~\min_\mu c^\mu(j)~~~,~~~c^\mu(j)=\sum_i s_i(j)(a_i^\mu-\rho b_i^\mu)
\end{equation}
\item If $c^{\mu_0}(j)\geq0$, then $\mathbf{s}(j)$ is a solution and exit;
\item Else, if $c^{\mu_0}<0$, update fluxes as follows:
\begin{equation}\label{update}
s_i(j+1)=\max\{0,s_i(j)+a_i^{\mu_0(j)}-\rho b_i^{\mu_0(j)}\}
\end{equation}
and return to (ii).
\end{enumerate}
If at time $j$ more than one reagent satisfies (\ref{muz}), a single $\mu_0$ can be chosen by picking one at random with uniform probability among all metabolites having the minimum value of $c^\mu$. In addition, at the end of each step it is possible to normalize the flux vector appropriately (e.g. $\sum_i s_i=N$). Iteration of this subroutine for increasing values of $\rho$ allows to compute $\rho^\star$ with the desired degree of accuracy. Note that convergence to a solution is guaranteed for every $\rho<\rho^\star$ and, in addition, it can be shown that the algorithm samples convex solution sets uniformly.

To extend this procedure to the reversible model it is not sufficient to straightforwardly eliminate the lower bound in (\ref{update}), since the reinforcement term in this case is not fixed (as for a standard perceptron) but depends on the flux itself. In particular it is given by
\begin{equation}
\Delta_i^\mu= s_i \theta(s_i) (a_i^\mu-\rho b_i^\mu)- s_i\theta(-s_i) (b_i^\mu-\rho a_i^\mu)
\end{equation}
The non-linearity of the above relation deforms the `patterns' each time the direction of a reaction is reversed, so it is not immediately obvious in which direction it is necessary to modify a flux to approach the solution. 

For reversible reactions, we thus introduce a different core algorithm to solve Von Neumann's conditions (to be iterated as before over $\rho$ to calculate $\rho^\star$). Let us set
\begin{equation}\fl
X_i^\mu(\rho)=a_i^\mu-\rho b_i^\mu~~~~~,~~~~~
Y_i^\mu(\rho)=b_i^\mu-\rho a_i^\mu~~~~~,~~~~~
h_i^\mu(\rho)=X_i^\mu(\rho)-Y_i^\mu(\rho)
\end{equation}
The first two quantities represent, respectively, the net amount of metabolite $\mu$ that the $i$-th reaction gives to (if positive) or subtracts from (if negative) the system per unit of flux, in straight and reverse direction respectively. The last one, equal to $(1+\rho)(a_i^\mu-b_i^\mu)$, has the same sign of direction that the $i$-th reaction has to take to produce the largest amount of metabolite $\mu$ per unit of flux. We proceed as follows.
\begin{enumerate}
\item Step $j=0$:  randomly initialize the flux vector $\mathbf{s}(0)$ (e.g. as a random vector with uniformly distributed entries in $[-1,1]$);
\item Step $j+1$: compute
\begin{equation}
\mu_0={\rm arg}~\min_\mu c^\mu(j)
\end{equation}
with
\begin{equation}
c^\mu(j)\equiv\sum_{i=1}^N s_i(j) [\theta[s_i(j)](a_i^\mu-\rho b_i^\mu)-\theta[-s_i(j)](b_i^\mu-\rho a_i^\mu)] 
\end{equation}
\item If $c^{\mu_0}(j)\geq0$, then $\mathbf{s}(j)$ is a solution and exit;;
\item Else, if $c^{\mu_0}(j)<0$, update fluxes as follows:
\begin{itemize}
\item[(a)] If $\min \{(X_i^{\mu_0}(\rho),Y_i^{\mu_0}(\rho)\} \leq 0$ then
\begin{equation}\fl
s_i(j+1)= 
\cases{
\max\{0,s_i(j)+a_i^{\mu_0(j)}-\rho b_i^{\mu_0(j)}\} & if $\quad s_i(j)>0$\\
\min\{0,s_i(j)-(b_i^{\mu_0(j)}-\rho a_i^{\mu_0(j)})\} & if $\quad s_i(j)<0$\\
h_i^\mu(\rho) & if $\quad s_i(j)=0$}
\end{equation}
\item[(b)] Else, if $\min\{(X_i^{\mu_0}(\rho),Y_i^{\mu_0}(\rho)\}>0$:
\begin{itemize}
\item[(b.1)] If  $s_i(j)h_i^{\mu_0}(\rho)\geq 0$ follow the same instructions as in (a);
\item[(b.2)] Else, if  $s_i(j)h_i^{\mu_0}(\rho)< 0$ update fluxes as
\begin{equation}
s_i(j+1)= s_i(j)+h_i^{\mu_0}(\rho) \quad
\end{equation}
and return to (ii).
\end{itemize}
\end{itemize}
\end{enumerate}

This procedure tries in essence to favor for each reaction the direction allowing for an increase of the production of the reagent corresponding to the least satisfied constraint, unless both directions are feasible (i.e. unless both $X_i^{\mu_0}$ and $Y_i^{\mu_0}$ are positive), in which case the flux vector is rotated in the most advantageous direction. Finally, if at the previous step the reaction was inactive, the reaction has to be activated in the direction that favors maximal production of $\mu_0$.

In a network with partial reversibility, it is possible to update each flux using the reversible or the irreversible algorithm according to the type of reaction considered. We have been unable to prove convergence (as was instead possible for the Minover$^+$ algorithm) and the only support to its effectiveness lies (besides numerous tests) in its ability to recover the replica prediction.

\end{document}